\documentclass[12pt,a4paper]{article}

\usepackage{lineno}  
\usepackage{graphicx}  
\usepackage{xspace} 
\usepackage{color}
\usepackage{colortbl}
\usepackage{amsmath} 
\usepackage{ifthen} 

\newboolean{pdflatex}
\setboolean{pdflatex}{true} 
\newboolean{articletitles}
\setboolean{articletitles}{false} 

\newboolean{uprightparticles}
\setboolean{uprightparticles}{false} 

\usepackage{amssymb}
\usepackage{amsfonts}
\usepackage{upgreek} 

\usepackage{hyperref}    
\usepackage[all]{hypcap} 

\def\ux85 {UX85\xspace}

\ifthenelse{\boolean{uprightparticles}}%
{

 \def\PDelta      {\ensuremath{\Delta}\xspace}                 
 \def\PXi      {\ensuremath{\Xi}\xspace}                 
 \def\PLambda      {\ensuremath{\Lambda}\xspace}                 
 \def\PSigma      {\ensuremath{\Sigma}\xspace}                 
 \def\POmega      {\ensuremath{\Omega}\xspace}                 
 \def\PUpsilon      {\ensuremath{\Upsilon}\xspace}                 
 
 \def\PB      {\ensuremath{\mathrm{B}}\xspace}                 
                  
 \def\PD      {\ensuremath{\mathrm{D}}\xspace}

 \def\PK      {\ensuremath{\mathrm{K}}\xspace}

 \def\Pi      {\ensuremath{\mathrm{i}}\xspace}

 \def\Ps      {\ensuremath{\mathrm{s}}\xspace}

}
{

 \mathchardef\PDelta="7101
 \mathchardef\PXi="7104
 \mathchardef\PLambda="7103
 \mathchardef\PSigma="7106
 \mathchardef\POmega="710A
 \mathchardef\PUpsilon="7107
                  
 \def\PB      {\ensuremath{B}\xspace}                 
                  
 \def\PD      {\ensuremath{D}\xspace}

 \def\PK      {\ensuremath{K}\xspace}

 \def\Pi      {\ensuremath{i}\xspace}

 \def\Ps      {\ensuremath{s}\xspace}

}

\def\squark    {\ensuremath{\Ps}\xspace}

\def\kaon  {\ensuremath{\PK}\xspace}
  \def\Kbar  {\kern 0.2em\overline{\kern -0.2em \PK}{}\xspace}

\def\Kz    {\ensuremath{\kaon^0}\xspace}
\def\Kzb   {\ensuremath{\Kbar^0}\xspace}
\def\KzKzb {\ensuremath{\Kz \kern -0.16em \Kzb}\xspace}
\def\Kp    {\ensuremath{\kaon^+}\xspace}
\def\Km    {\ensuremath{\kaon^-}\xspace}

\def\KpKm  {\ensuremath{\Kp \kern -0.16em \Km}\xspace}

\def\Kstarzb {\ensuremath{\Kbar^{*0}}\xspace}

  \def\Dbar    {\kern 0.2em\overline{\kern -0.2em \PD}{}\xspace}
\def\D       {\ensuremath{\PD}\xspace}

\def\Dz      {\ensuremath{\D^0}\xspace}
\def\Dzb     {\ensuremath{\Dbar^0}\xspace}
\def\DzDzb   {\ensuremath{\Dz {\kern -0.16em \Dzb}}\xspace}
\def\Dp      {\ensuremath{\D^+}\xspace}
\def\Dm      {\ensuremath{\D^-}\xspace}

\def\DpDm    {\ensuremath{\Dp {\kern -0.16em \Dm}}\xspace}

\def\B       {\ensuremath{\PB}\xspace}
  \def\Bbar    {\kern 0.18em\overline{\kern -0.18em \PB}{}\xspace}

\def\Bzb     {\ensuremath{\Bbar^0}\xspace}

\def\Bs      {\ensuremath{\B^0_\squark}\xspace}
\def\Bsb     {\ensuremath{\Bbar^0_\squark}\xspace}

  \def\Y#1S{\ensuremath{\PUpsilon{(#1S)}}\xspace}

\def\Lb      {\ensuremath{\L_\bquark}\xspace}

\def\Lc      {\ensuremath{\L_\cquark}\xspace}

\def\to                 {\ensuremath{\rightarrow}\xspace}

\newcommand{\dms}{\ensuremath{\Delta m_{\squark}}\xspace}

\def\AT#1     {\ensuremath{A_{\mathrm{T}}^{#1}}\xspace}           

\def\C#1      {\ensuremath{\mathcal{C}_{#1}}\xspace}                       
\def\Cp#1     {\ensuremath{\mathcal{C}_{#1}^{'}}\xspace}                    
\def\Ceff#1   {\ensuremath{\mathcal{C}_{#1}^{\mathrm{(eff)}}}\xspace}        
\def\Cpeff#1  {\ensuremath{\mathcal{C}_{#1}^{'\mathrm{(eff)}}}\xspace}       
\def\Ope#1    {\ensuremath{\mathcal{O}_{#1}}\xspace}                       
\def\Opep#1   {\ensuremath{\mathcal{O}_{#1}^{'}}\xspace}                    

\newcommand{\tev}{\ensuremath{\mathrm{\,Te\kern -0.1em V}}\xspace}
\newcommand{\gev}{\ensuremath{\mathrm{\,Ge\kern -0.1em V}}\xspace}
\newcommand{\mev}{\ensuremath{\mathrm{\,Me\kern -0.1em V}}\xspace}
\newcommand{\kev}{\ensuremath{\mathrm{\,ke\kern -0.1em V}}\xspace}
\newcommand{\ev}{\ensuremath{\mathrm{\,e\kern -0.1em V}}\xspace}
\newcommand{\gevc}{\ensuremath{{\mathrm{\,Ge\kern -0.1em V\!/}c}}\xspace}
\newcommand{\mevc}{\ensuremath{{\mathrm{\,Me\kern -0.1em V\!/}c}}\xspace}
\newcommand{\gevcc}{\ensuremath{{\mathrm{\,Ge\kern -0.1em V\!/}c^2}}\xspace}
\newcommand{\gevgevcccc}{\ensuremath{{\mathrm{\,Ge\kern -0.1em V^2\!/}c^4}}\xspace}
\newcommand{\mevcc}{\ensuremath{{\mathrm{\,Me\kern -0.1em V\!/}c^2}}\xspace}

\def\mum  {\ensuremath{\,\upmu\rm m}\xspace}

\def\gsim{{~\raise.15em\hbox{$>$}\kern-.85em
          \lower.35em\hbox{$\sim$}~}\xspace}
\def\lsim{{~\raise.15em\hbox{$<$}\kern-.85em
          \lower.35em\hbox{$\sim$}~}\xspace}

\def\tell1  {TELL1\xspace}
\def\ukl1   {UKL1\xspace}

\usepackage{mciteplus}

\begin{document}

\def\dzdzb{{~\raise.85em\hbox{{\tiny{(}\textemdash\tiny{)}}}\kern-1.05em
          \lower0.0em\hbox{$D^0$}~}\xspace}
\def\bsbsb{{~\raise.85em\hbox{{\tiny{(}\textemdash\tiny{)}}}\kern-1.05em
          \lower0.0em\hbox{$B_s^0$}~}\xspace}

\mathchardef\mhyphen="2D

\def\nsig{N_{\rm sig}}
\def\nback{N_{\rm back}}
\def\bb{b\bar{b}}
\def\bs{B_s}
\def\bsb{\bar{B}_s}
\def\dsp{D_s^+}
\def\dsm{D_s^-}
\def\dspm{D_s^{\pm}}
\def\lam{\lambda}
\def\lamp{\lambda^{\prime}}
\def\bsdsk{\bsbsb\to D_s^{\mp} K^{\pm}}
\def\bsdsstark{B_s\to D_s^{\mp*} K^{\pm}}
\def\bsdskst{B_s\to D_s^{\mp} K^{*\pm}, K^{*\pm}\to K_s\pi^+}
\def\bsdskpi{B_s\to D_s^{\mp} K^{*\pm}, K^{*\pm}\to K^+\pi^0}
\def\bsdsp{B_s\to D_s^{\mp} \pi^{\pm}}
\def\bsdskpp{B_s\to D_s^{\mp} K^{\pm}\pi^{\mp}\pi^{\pm}}
\def\bsdskppa{B_s\to D_s^{-} (K^+\pi^-\pi^+)}
\def\bsdskppa{\bar{B}_s\to D_s^{-} (K^+\pi^-\pi^+)}
\def\bsbdsk{B_s\to D_s^{\pm} K^{\mp}}
\def\dg{\Delta\Gamma_s}
\def\dms{\Delta m_s}
\def\dd{{\cal{D}}}
\def\br{{\cal{B}}}

\def\xb{H_b}
\def\xc{H_c}
\def\Lb{\Lambda_b^0}
\def\Lc{\Lambda_c^+}
\def\Sc{\Sigma_c}

\def\pimm{\pi{\raisebox{2.5pt}{$\scriptstyle-$}}{\mspace{-11mu}\raisebox{5.25pt}{$\scriptstyle-$}}}

\def\xbtoxchhh{\xb\to\xc h^-h^+h^-}
\def\xbtoxcpipipi{\xb\to\xc\pi^-\pi^+\pi^-}
\def\xbtoxcpi{\xb\to \xc\pi^-}
\def\xbtoxch{\xb\to \xc h^-}
\def\xbtoxckpipi{\xb\to\xc K^-\pi^+\pi^-}
\def\xbtoxck{\xb\to\xc K^-}

\def\efake{\epsilon_{misID}(p,p_T,\eta)}

\def\btodhhh{B\to D + 3h}
\def\bstodspipipi{\Bs\to D_s^+\pi^-\pi^+\pi^-}
\def\bstodskpipi{\Bs\to D_s^{\mp}K^{\pm}\pi^+\pi^-}
\def\btodpipipi{\Bzb\to D^+\pi^-\pi^+\pi^-}
\def\btodkpipi{\Bzb\to D^+K^-\pi^+\pi^-}
\def\btodstarkpipi{\Bzb\to D^{*+}K^-\pi^+\pi^-}
\def\btodzeropipipi{B^-\to D^0\pi^-\pi^+\pi^-}
\def\btodzerokpipi{B^-\to D^0 K^-\pi^+\pi^-}
\def\btodstarzerokpipi{B^-\to D^{*0} K^-\pi^+\pi^-}

\def\btodstarpipipi{\Bzb\to D^{*+}\pi^-\pi^+\pi^-}
\def\LbtoLcpipipi{\Lambda_b^0\to\Lambda_c^+\pi^-\pi^+\pi^-}
\def\LbtoLckpipi{\Lambda_b^0\to\Lambda_c^+ K^-\pi^+\pi^-}
\def\LbtoLcpi{\Lambda_b^0\to\Lambda_c^+\pi^-}
\def\LbtoLck{\Lambda_b^0\to\Lambda_c^+K^-}
\def\btodh{B\to Dh}
\def\bstodspi{\Bsb\to D_s^+\pi^-}
\def\bstodsk{\Bs\to D_s^{\mp}K^{\pm}}
\def\btodpi{\Bzb\to D^+\pi^-}
\def\btodk{\Bzb\to D^+K^-}
\def\btodstarpi{\Bzb\to D^{*+}\pi^-}
\def\btodzeropi{B^{-}\to D^0\pi^{-}}
\def\btodzerok{B^-\to D^0 K^-}
\def\btodzerobark{B^-\to D K^-}
\def\eff{\epsilon}
\def\btodzerokstar{\Bzb\to D^0\Kstarzb}

\def\ifb{\rm fb^{-1}}
\def\ipb{\rm pb^{-1}}



\begin{titlepage}
\pagenumbering{roman}

\vspace*{-1.5cm}
\centerline{\large EUROPEAN ORGANIZATION FOR NUCLEAR RESEARCH (CERN)}
\vspace*{1.5cm}
\hspace*{-0.5cm}
\begin{tabular*}{\linewidth}{lc@{\extracolsep{\fill}}r}
\ifthenelse{\boolean{pdflatex}}
{\vspace*{-2.7cm}\mbox{\!\!\!\includegraphics[width=.14\textwidth]{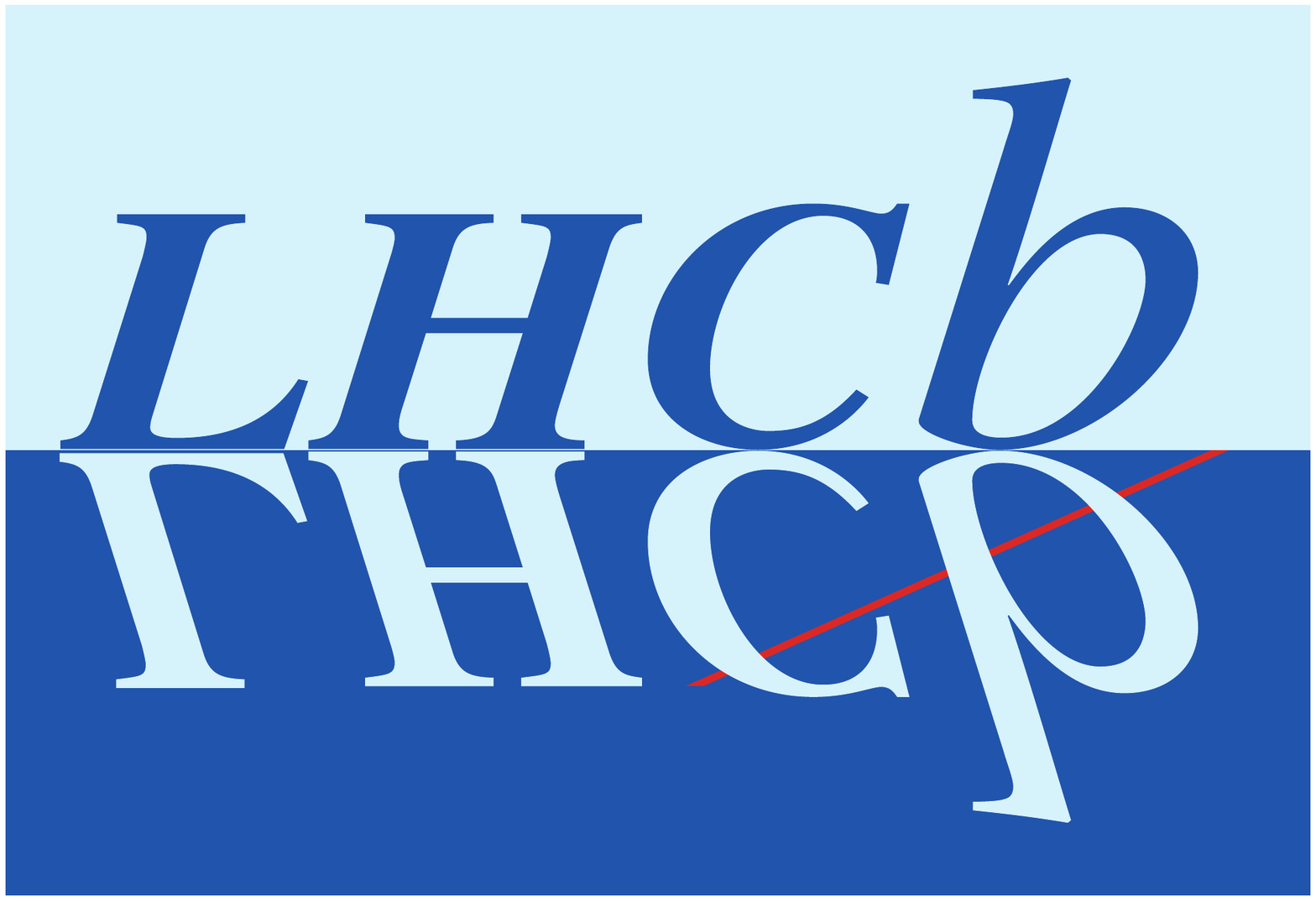}} & &}%
{\vspace*{-1.2cm}\mbox{\!\!\!\includegraphics[width=.12\textwidth]{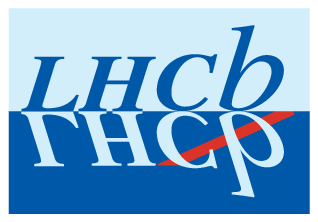}} & &}%
\\
 & & CERN-PH-EP-2011-229 \\  
 & & LHCb-PAPER-2011-040 \\  
 & & \today \\ 
 & & \\
\end{tabular*}

\vspace*{4.0cm}

{\bf\boldmath\huge
\begin{center}
  First observation of the decays $\btodkpipi$ and $\btodzerokpipi$
\end{center}
}

\vspace*{2.0cm}

\begin{center}
The LHCb Collaboration
\footnote{Authors are listed on the following pages.}
\end{center}

\vspace{\fill}

\begin{abstract}
  \noindent
First observations of the Cabibbo suppressed decays $\btodkpipi$ and $\btodzerokpipi$ are reported using 35~$\ipb$ of
data collected with the LHCb detector. Their branching fractions are measured with respect to the 
corresponding Cabibbo favored decays, from which we obtain $\br(\btodkpipi)/\br(\btodpipipi)=(5.9\pm1.1\pm0.5)\times10^{-2}$
and $\br(\btodzerokpipi)/\br(\btodzeropipipi) = (9.4\pm 1.3\pm0.9)\times10^{-2}$, where the uncertainties
are statistical and systematic, respectively. The $\btodzerokpipi$ decay is particularly interesting, as it
can be used in a similar way to $\btodzerok$ to measure the CKM phase $\gamma$. 
\end{abstract}

\vspace*{1.0cm}
\begin{center}
To be submitted to Physical Review Letters.
\end{center}
\vspace{\fill}

\end{titlepage}




\centerline{\large\bf The LHCb Collaboration}
\begin{flushleft}
\small
R.~Aaij$^{23}$, 
C.~Abellan~Beteta$^{35,n}$, 
B.~Adeva$^{36}$, 
M.~Adinolfi$^{42}$, 
C.~Adrover$^{6}$, 
A.~Affolder$^{48}$, 
Z.~Ajaltouni$^{5}$, 
J.~Albrecht$^{37}$, 
F.~Alessio$^{37}$, 
M.~Alexander$^{47}$
% ...

\noindent {\footnotesize \it
% ...
$ ^{5}$Clermont Universit\'{e}, Universit\'{e} Blaise Pascal, CNRS/IN2P3, LPC, Clermont-Ferrand, France\\
$ ^{6}$CPPM, Aix-Marseille Universit\'{e}, CNRS/IN2P3, Marseille, France\\
% ...
$ ^{23}$Nikhef National Institute for Subatomic Physics, Amsterdam, The Netherlands\\
% ...
$ ^{35}$Universitat de Barcelona, Barcelona, Spain\\
$ ^{36}$Universidad de Santiago de Compostela, Santiago de Compostela, Spain\\
$ ^{37}$European Organization for Nuclear Research (CERN), Geneva, Switzerland\\
% ...
$ ^{42}$H.H. Wills Physics Laboratory, University of Bristol, Bristol, United Kingdom\\
% ...
$ ^{47}$School of Physics and Astronomy, University of Glasgow, Glasgow, United Kingdom\\
$ ^{48}$Oliver Lodge Laboratory, University of Liverpool, Liverpool, United Kingdom\\
% ...
$ ^{n}$LIFAELS, La Salle, Universitat Ramon Llull, Barcelona, Spain\\
% ...
}
\bigskip
\end{flushleft}

\cleardoublepage




\pagestyle{plain} 
\setcounter{page}{1}
\pagenumbering{arabic}


%

\noindent 
The standard model (SM) of particle physics provides a good description of nature up to the TeV
scale, yet many issues remain unresolved~\cite{Quigg:2009vq}, including, but not limited to, the hierarchy problem, 
the preponderance of matter over antimatter in the Universe, and the need to explain dark matter.
One of the main objectives of the LHC is to search for new physics beyond the Standard Model (SM) either through 
direct detection or through interference effects in $b$- and $c$-hadron decays. 
In the SM, the Cabibbo-Kobayashi-Maskawa (CKM) matrix~\cite{Cabibbo:1963yz,*Kobayashi:1973fv}
governs the strengths of weak charged-current interactions and their corresponding phases. 
Precise measurements on the CKM matrix parameters may reveal deviations from the consistency that is 
expected in the SM, making study of these decays a unique laboratory in which to search for physics
beyond the standard model. 

The most poorly constrained of the CKM parameters is the weak phase 
$\gamma\equiv {\rm arg}\left(-{V_{\rm ub}^*V_{\rm ud}\over V_{\rm cb}^*V_{\rm cd}}\right)$.  
Its direct measurement reaches a precision of $10^{\circ}-12^{\circ}$~\cite{Bevan:2010zz,Charles:2011va}.
Two promising methods of
measuring this phase are through the time-independent and time-dependent analyses of 
$\btodzerok$~\cite{Dunietz:1991yd,*Dunietz:1992ti,*Atwood:1994zm,*Atwood:1996ci,Gronau:1990ra,*Gronau:1991dp,Giri:2003ty}
and $\bstodsk$~\cite{Aleksan:1991nh,Dunietz:1995cp}, respectively. Both approaches
can be extended to higher multiplicity modes, such as $\btodzerokstar$, $\btodzerokpipi$~\cite{Gronau:2002mu} 
and $\bstodskpipi$, which could provide a comparable level of sensitivity. The last
two decays have not previously been observed. 

In this Letter, we report first observations of the Cabibbo-suppressed (CS) $\btodkpipi$
and $\btodzerokpipi$ decays, where $D^+\to K^-\pi^+\pi^+$ and $D^0\to K^-\pi^+$, where
charge conjugation is implied throughout this Letter.
These signal decays are normalized with respect to the topologically similar
Cabibbo-favored (CF) $\btodpipipi$ and $\btodzeropipipi$ decays, respectively. For brevity, 
we use the notation $X_d$ to refer to the recoiling $\pi^-\pi^+\pi^-$ system in the CF decays
and $X_s$ for the $K^-\pi^+\pi^-$ system in the CS decays.

The analysis presented here is based on 35~$\ipb$ of data collected with the LHCb detector in
2010. For these measurements, the most important parts of LHCb are the
vertex detector (VELO), the charged particle tracking system, the ring imaging Cherenkov detectors 
(RICH) and the trigger.
The VELO is instrumental in separating particles coming from heavy quark decays
and those emerging directly from $pp$ interactions,  by providing an impact parameter (IP) resolution
of about $16\mum$ + 30$\mum$/$p_{\rm T}$ (transverse momentum, $p_{\rm T}$ in GeV/$c$). The tracking system measures 
charged particles' momenta with a resolution of $\sigma_p/p\sim0.4\% (0.6\%)$ at 5 (100) GeV/$c$. 
The RICH detectors are important to identify kaons and suppress the large backgrounds from pions 
misidentified as kaons. Events are selected by a two-level 
trigger system. The first level is 
hardware-based, and requires either a large transverse energy deposition in the calorimeter system,
or a high $p_{\rm T}$ muon or pair of muons detected in the muon system. 
The second level, the high-level trigger, 
uses simplified versions of the offline software to reconstruct decays of $b$- and $c$-hadrons
both inclusively and exclusively. Candidates passing the trigger selections are
saved and used for offline analysis. 
A more detailed description of the LHCb detector can be found elsewhere~\cite{Alves:2008zz}.
In this analysis the signal and normalization modes are topologically identical, 
allowing loose trigger requirements to be made with small associated uncertainty.
In particular, we exploit the fact that $b$-hadrons are produced in pairs in $pp$ collisions, and include 
events that were triggered by the decay products of either the signal $b$-hadron or the other $b$-hadron in the 
event. This requirement increases the efficiency of our trigger selection by about 80\%
compared to the trigger selections requiring the signal $b$-hadron to be responsible
for triggering the event, as was done in Ref.~\cite{Aaij:2011rj}.

The selection criteria used to reconstruct the $\btodpipipi$ and
$\btodzeropipipi$ final states are described in Ref.~\cite{Aaij:2011rj}.
The Cabibbo suppression results in about a factor of 20 lower rate.
To improve the signal-to-background ratio in the CS decay modes, additional selection
requirements are imposed, and they are applied to both the signal and normalization modes. 
The $B$ meson candidate is required to have $p_{\rm T}>4$~GeV/$c$, ${\rm IP}<60~\mum$ with respect to its associated 
primary vertex (PV), where the associated PV is the one having the smallest impact parameter 
$\chi^2$ with respect to the track. We also require the flight distance $\chi^2>144$,
where the $\chi^2$ is with respect to the zero flight distance hypothesis, and the vertex $\chi^2/{\rm ndf}<5$,
where ndf represents the number of degrees of freedom in the fit. The last requirement is also applied to
the vertices associated with $X_d$ and $X_s$. Three additional criteria are applied only to the
CS modes. First, to remove the peaking backgrounds from $B\to DD_s^-,~D_s^-\to K^-\pi^+\pi^-$,
we veto events where the invariant mass, $M(X_s)$, is within 20~MeV/$c^2$ of the $D_s$ mass. 
Information from the RICH is critical to reduce 
background from the CF decay modes. This suppression is accomplished by requiring the kaon in 
$X_s$ to have $p<100$~GeV/$c$ (above which there is minimal $K$/$\pi$ separation from the RICH),
and the difference in log-likelihoods between the kaon and pion hypotheses to
satisfy $\Delta{\rm ln}\mathcal{L}(K-\pi)>8$. The latter requirement is determined by optimizing $N_S/\sqrt{N_S+N_B}$,
where we assume 100 signal events ($\sim1/20$ of the CF decay yields) prior to 
any particle identification (PID) selection requirement, and the combinatorial 
background yield, $N_B$, is taken from the high $B$-mass sideband (5350-5580~MeV/$c^2$). We also make a
loose PID requirement of $\Delta{\rm ln}\mathcal{L}(K-\pi)<10$ on the pions in $X_s$ and $X_d$.

Selection and trigger efficiencies are determined from simulation. Events are produced using 
{\sc pythia}~\cite{Sjostrand:2006za} and long-lived particles are decayed using {\sc evtgen}~\cite{Lange:2001uf}.
The detector response is simulated with {\sc geant4}~\cite{Agostinelli:2002hh}. The $DK^-\pi^+\pi^-$ final states
are assumed to include 50\% $DK_1(1270)^-$ and 20\% $DK_1(1400)^-$,
with smaller contributions from $DK_2(1430)^-$, $DK^*(1680)^-$, $D\bar{K}^*(892)^0\pi^-$ and $D_1(2420)K^-$.
The resonances included in the simulation of the $X_d$ system are described in Ref.~\cite{Aaij:2011rj}. 
The relative efficiencies, including selection and trigger, but not PID selection, are determined to be
$\eff_{\btodkpipi}/\eff_{\btodpipipi} = 1.05\pm0.04$ and 
$\eff_{\btodzerokpipi}/\eff_{\btodzeropipipi} = 0.942\pm0.036$, where the uncertainties are statistical only.
The efficiencies have a small dependence on the contributing
resonances and their daughters' masses, and we therefore do not necessarily expect the ratios to be
equal to unity. Moreover, the additional selections on the CS modes contribute to small differences
between the signal and normalization modes' efficiencies.

The PID efficiencies are determined in bins of track momentum and pseudorapidity ($\eta$)
using the $D^0$ daughters from $D^{*\pm}\to\pi_s^{\pm}D^0$, $D^0\to K^-\pi^+$ calibration data, where the particles 
are identified without RICH information using the charge of the soft pion, $\pi_s$. The kinematics of the 
kaon in the $X_s$ system are taken 
from simulation after all offline and trigger selections. Applying the PID efficiencies to the 
simulated decays, we determine the efficiencies for the kaon to pass the $\Delta{\rm ln}\mathcal{L}(K-\pi)>8$
requirement to be $(75.9\pm1.5)\%$ for $\btodkpipi$ and $(79.2\pm1.5)\%$ for $\btodzerokpipi$.

Invariant mass distributions for the normalization and signal modes 
are shown in Fig.~\ref{fig:b2da1cs}. Signal yields are
determined through unbinned maximum likelihood fits to the sum of signal and several
background components. The signal distributions are
parametrized as the sum of two Gaussian functions with common means, and
shape parameters,  $\sigma_{\rm core}$ and $f_{\rm core}$ that represent the width
and area fraction of the narrower (core) Gaussian portion, and 
$r_w\equiv\sigma_{\rm wide}/\sigma_{\rm core}$, which is the ratio of the wider to narrower
Gaussian width.

\begin{figure}[h]
\centering
\includegraphics[width=0.45\textwidth]{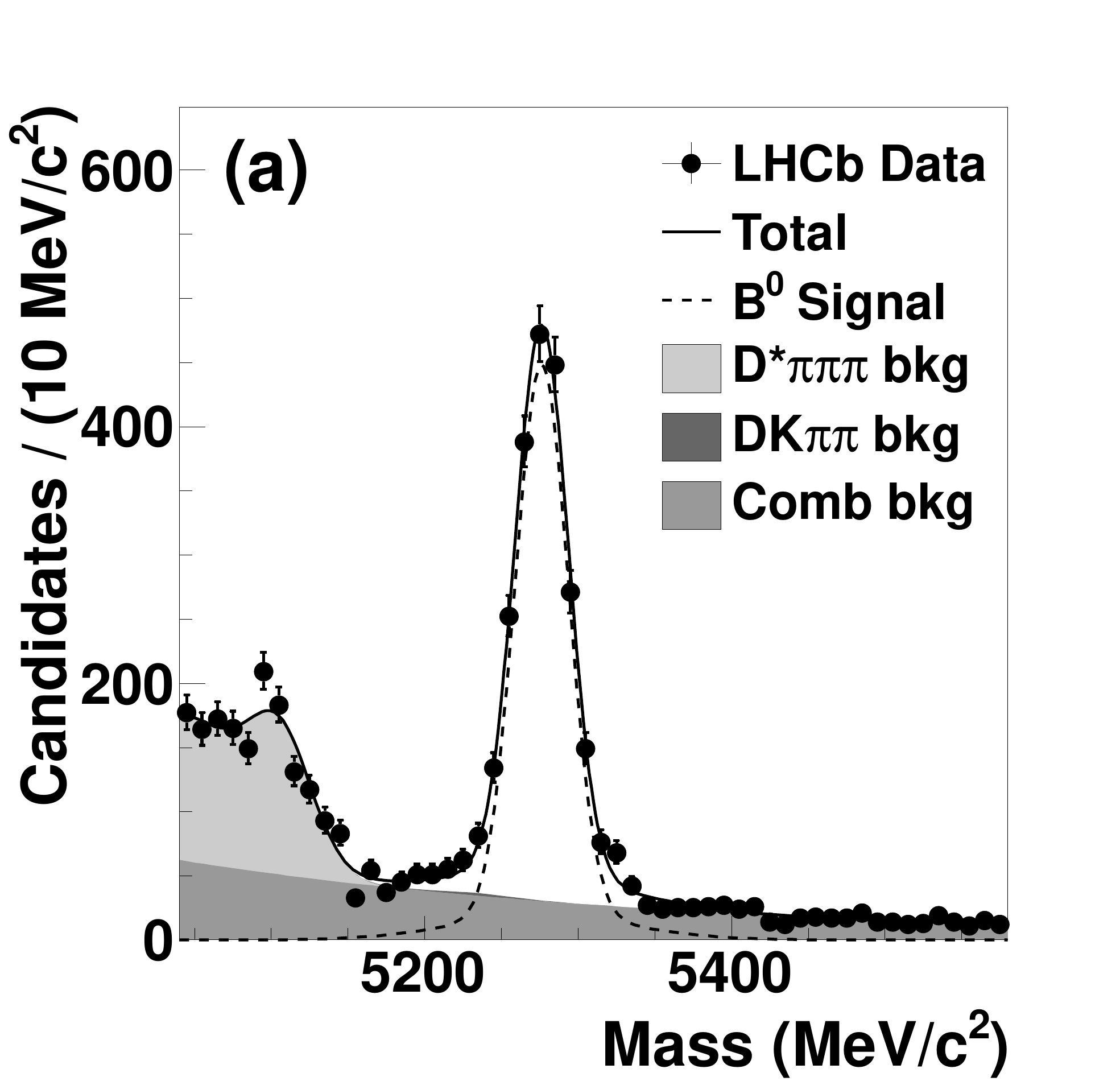}
\includegraphics[width=0.45\textwidth]{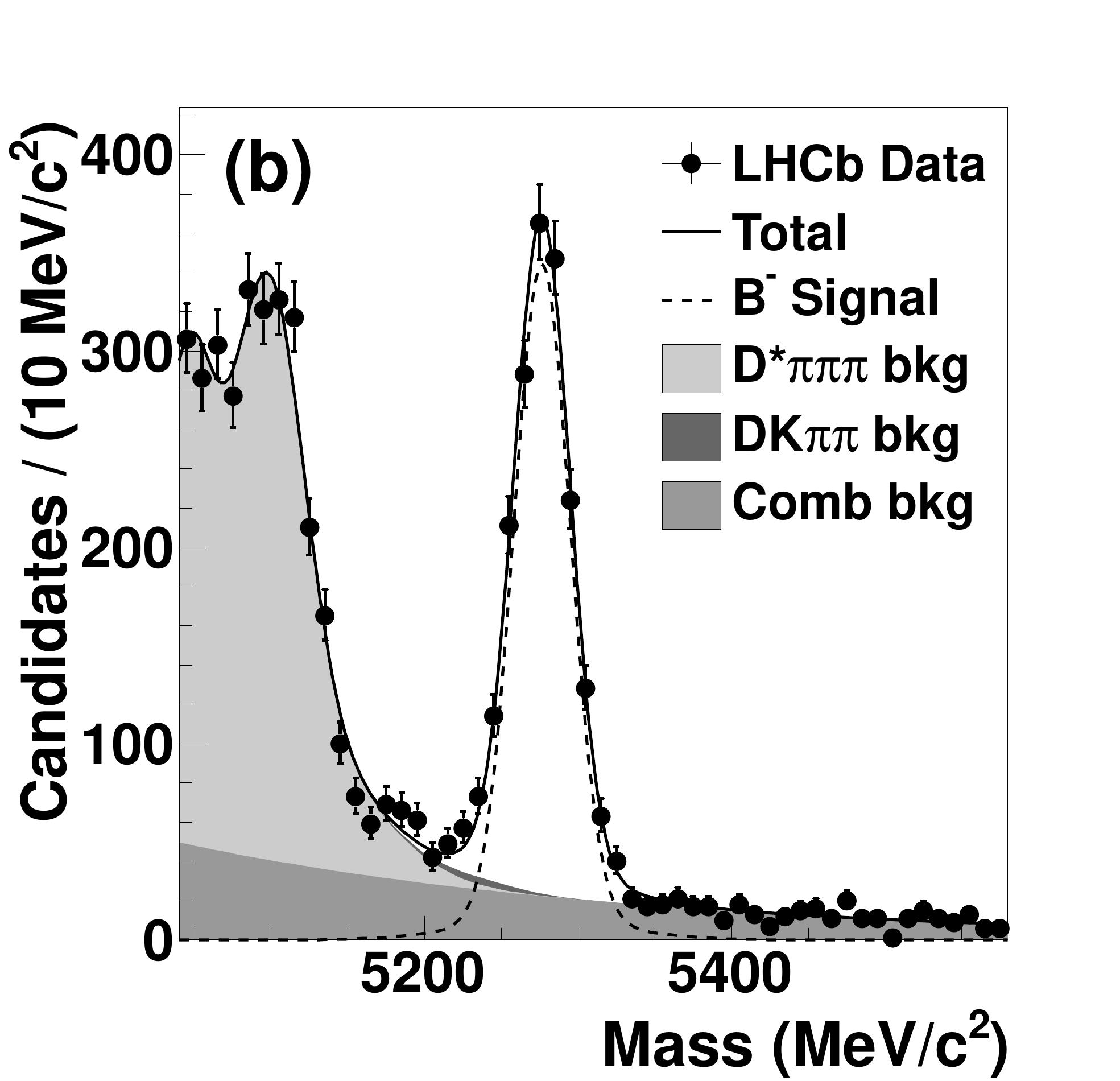}
\includegraphics[width=0.45\textwidth]{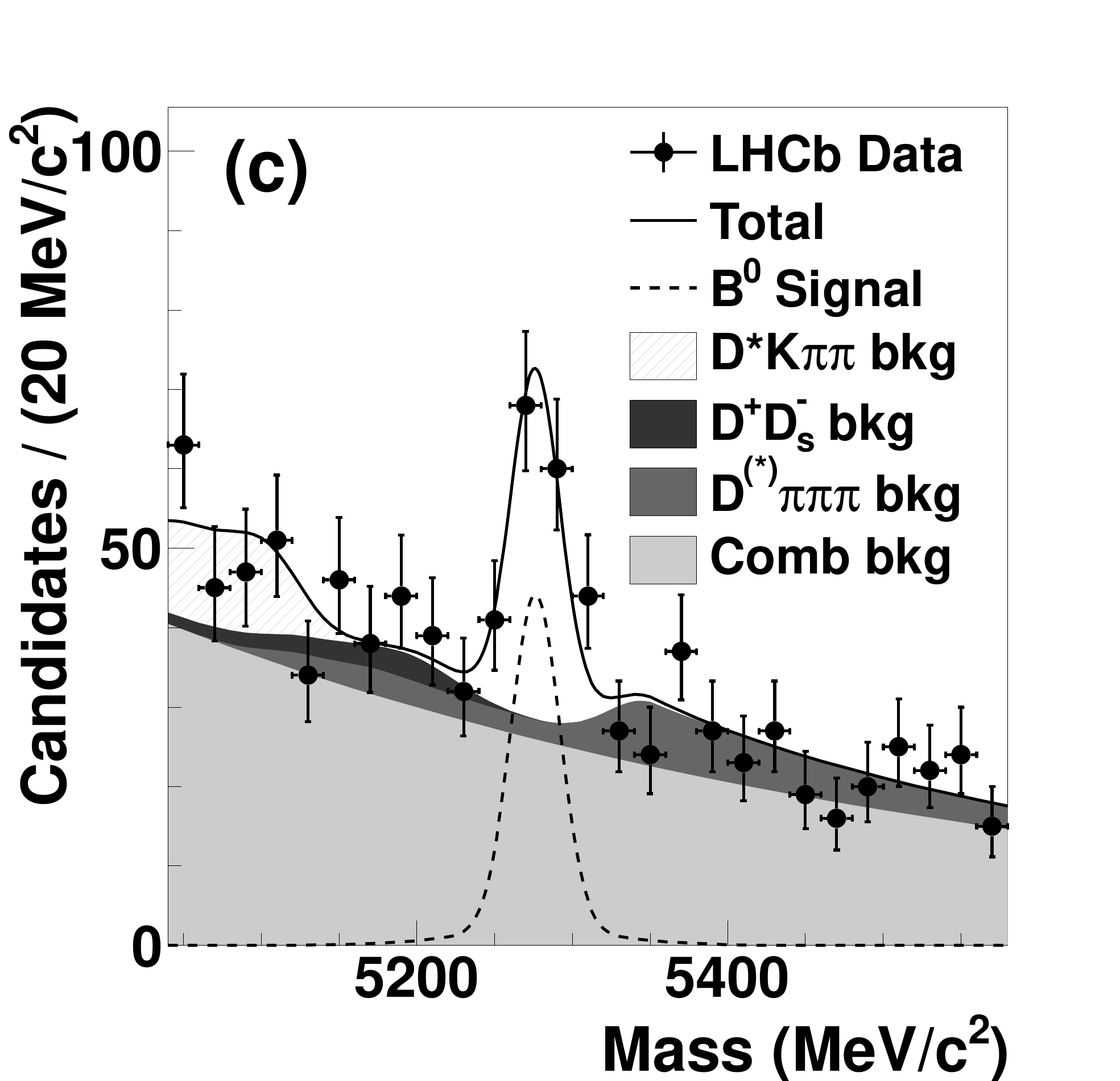}
\includegraphics[width=0.45\textwidth]{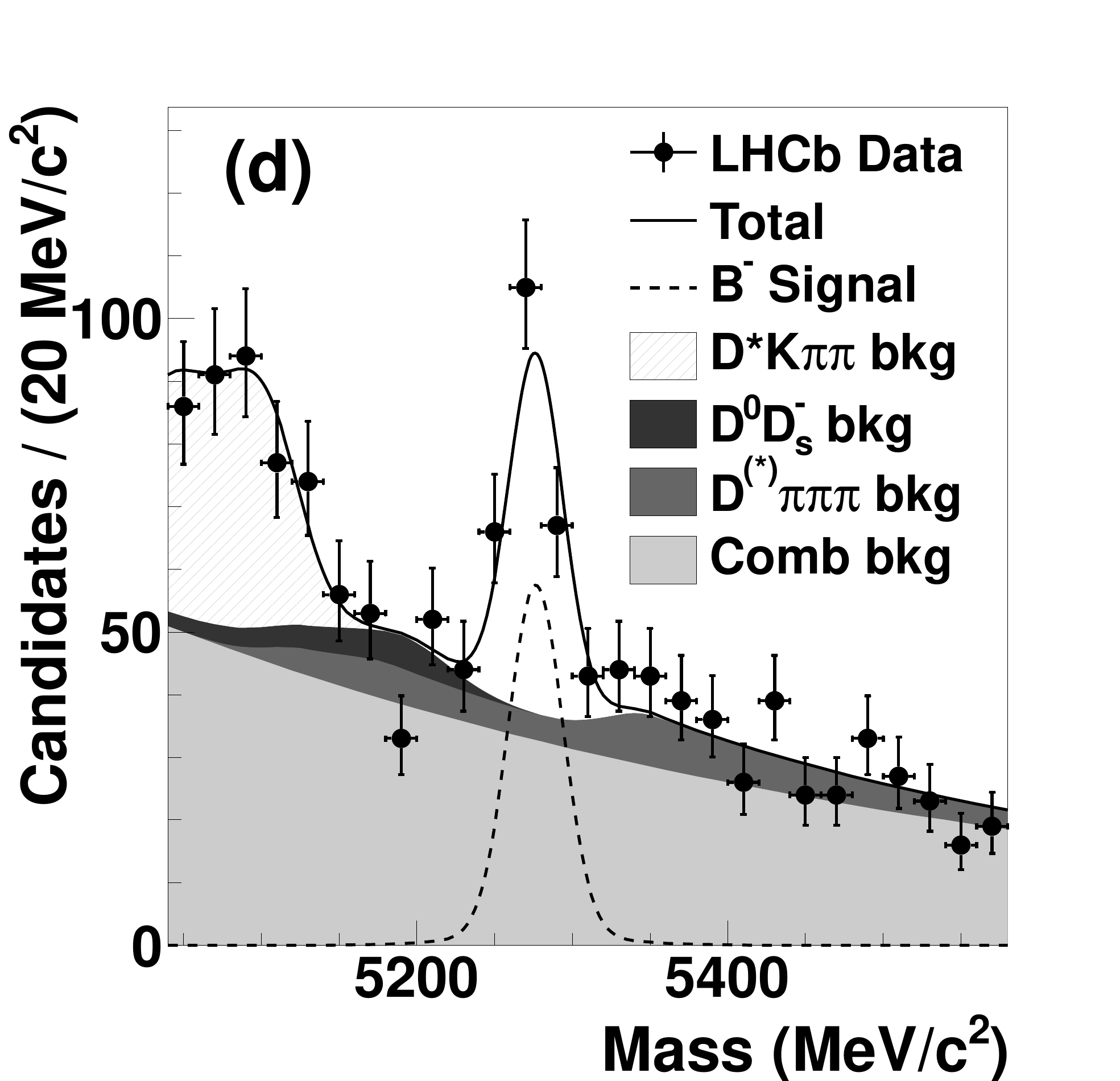}
\caption{Invariant mass distributions for (a) $\btodpipipi$, (b) $\btodzeropipipi$,
(c) $\btodkpipi$ and (d) $\btodzerokpipi$ candidates from 35~$\ipb$ of data for 
all selected candidates. Fits as described in the text are overlaid.}
\label{fig:b2da1cs}
\end{figure}

The CF modes are first fit with $f_{\rm core}$ and $r_w$ constrained to the values from
simulation within their uncertainties, while $\sigma_{\rm core}$ is left as a free parameter 
since simulation underestimates the mass resolution by $\sim$10\%. 
For the CF decay mode fits, the background shapes are the same as those described in 
Ref.~\cite{Aaij:2011rj}. 
The resulting signal shape parameters from the CF decay fits are then fixed in subsequent fits to 
the CS decay modes. For $\sigma_{\rm core}$, the values from the CF decay fits are scaled
by width correction factors ($\sim$0.95) obtained from MC simulations. 

For the CS decays, invariant mass shapes of specific peaking backgrounds 
from other $b$-hadron decays are determined from MC simulation. 
The largest of these backgrounds comes from $D^{(*)}\pi^-\pi^+\pi^-$ decays, where one of the 
$\pi^-$ passes the $\Delta{\rm ln}\mathcal{L}(K-\pi)>8$ requirement and is misidentified as a $K^-$. To
determine the fraction of events in which this occurs, we use measured PID fake rates ($\pi$ faking $K$) 
obtained from $D^{*\pm}$ calibration data (binned in $(p, \eta$)), and apply them to each $\pi^-$ in 
$D\pi^-\pi^+\pi^-$ simulated events. A decay is considered a fake if either pion has 
$p<100$~GeV/$c$, and a randomly generated number in the interval from [0, 1] is less than that 
pion's determined fake rate. The pion's mass is then replaced by the kaon's mass, and the 
invariant mass of the $b$-hadron is recomputed. The resulting spectrum is
then fitted using a Crystal Ball~\cite{Skwarnicki:1986xj} lineshape and its parameters are fixed in fits to data.
Using this method, we find the same cross-feed rate of $(4.4\pm0.7)\%$  for both
$\btodpipipi$ and $\btodzeropipipi$ into $\btodkpipi$ and $\btodzerokpipi$, respectively,
where the uncertainty includes both statistical and systematic sources.
A similar procedure is used to obtain the $D^*\pi^-\pi^+\pi^-$ background yields and shapes.
The background yields are obtained by multiplying the observed CF signal yields in data by the cross-feed
rates and the fraction of background in the region of the mass fit (5040$-$5580~MeV/$c^2$).

We also account for backgrounds from the decays $B\to DD_s^-,~D_s^-\to K^-K^+\pi^-$, where the $K^+$ is
misidentified as a $\pi^+$. The yields of these decays are lower, but are offset by a larger fake rate
since the PID requirement on the particles assumed to be pions is significantly looser ($\Delta{\rm ln}\mathcal{L}(K-\pi)<10$). 
Using the same technique as described above, the fake rate is found to be $(24\pm2)\%$. The
fake yield from this source is then computed from the product of the measured yield of $\Bzb\to D^+D_s^-$ in 
data ($161\pm14({\rm stat})$) and the above fake rate.
The $B^-\to D^0D_s^-$ yield was not directly measured, but was determined
from known branching fractions~\cite{Nakamura:2010zzi} and efficiencies from simulation. Additional uncertainty
due to these extrapolations is included in the estimated $B^-\to D^0D_s^-$ background yield.

The last sources of background, which do not contribute to the signal regions, are from $D^{*}K^-\pi^+\pi^-$,
where the soft pion or photon from the $D^*$ is lost. The
shapes of these low mass backgrounds are taken from the fitted $D^*\pi^-\pi^+\pi^-$ shapes in the 
$D\pi^-\pi^+\pi^-$ mass fits, and the yield ratios
$N(D^*K^-\pi^+\pi^-)/N(DK^-\pi^+\pi^-)$, are constrained 
to be equal to the ratios obtained from CF mode fits with a 25\% uncertainty.

The combinatorial background is assumed to have an exponential shape.
A summary of the signal shape parameters and the specific $b$-hadron backgrounds used in the
CS signal mode fits is given in Table~\ref{tab:csmodepar}.

The fitted yields are $2126\pm69$ $\btodpipipi$ and $1630\pm57$ and $\btodzeropipipi$ events.
For the CS modes, we find $90\pm16$ $\btodkpipi$ and $130\pm17$ $\btodzerokpipi$
signal decays. The CS decay signals have significances of 7.2 and 9.0, respectively, 
calculated as $\sqrt{-2{\rm ln}({\mathcal{L}_{0}}/{\mathcal{L}_{\rm max}}})$,
where ${\mathcal{L}_{\rm max}}$ and ${\mathcal{L}_{0}}$ are the fit likelihoods with the signal yields 
left free and fixed to zero, respectively.  
In evaluating these significances, we remove the constraint on $N(D^*K^-\pi^+\pi^-)/N(DK^-\pi^+\pi^-)$,
which would otherwise bias the $D^*K^-\pi^+\pi^-$ yield toward zero and inflate 
${\mathcal{L}_{0}}$. Varying the signal or background shapes 
or normalizations within their uncertainties has only a minor impact on the significances. We 
therefore observe for the first time the $\btodkpipi$ and $\btodzerokpipi$ decay modes.

\begin{table}[h]
\begin{center}
\caption{Summary of parameters used in the CS mass fits. Values without uncertainties are fixed in the
CS mode fits, and values with uncertainties are included with a Gaussian constraint with
central values and widths as indicated.}
\begin{tabular}{lcc}
\hline\hline \\ [-1.75ex]
Parameter               &    $~D^+K^-\pi^+\pi^-~$       &  $~D^0K^-\pi^+\pi^-~$ \\  
\hline\\ [-1.5ex]
Mean mass (MeV/$c^2$)           &       $5276.3$        &     $5276.5$ \\  
$\sigma_{\rm core}$  (MeV/$c^2$)   &       $15.7$        &     $17.5$ \\  
$f_{\rm core}$                     &    0.88   & 0.93 \\
$\sigma_{\rm wide}/\sigma_{\rm core}$ &  3.32 & 2.82 \\
$N(D\pi\pi\pi)$                 & $63\pm10$  & $48\pm8$ \\
$N(D^*\pi\pi\pi)$                 & $47\pm9~$  & $107\pm18$ \\
$N(DD_s)$                         & $23\pm3~$  & $38\pm8$ \\
$N(D^*K\pi\pi)/N(DK\pi\pi)$       & $0.62\pm0.16$ & $~1.86\pm0.46$ \\
\hline\hline
\end{tabular}
\label{tab:csmodepar}
\end{center}
\end{table}

The ratios of branching fractions are given by

\begin{equation}
  {\br(\xb\to\xc K^-\pi^+\pi^-)\over \br(\xb\to\xc\pi^-\pi^+\pi^-) } = {Y^{\rm CS}\over Y^{\rm CF} }\times\eff_{\rm tot}^{\rm rel},  \nonumber
\label{eq:bfeq}
\end{equation}

\noindent where $Y^{\rm CF}$ ($Y^{\rm CS}$) are the fitted yields in the CF (CS) decay modes, 
and $\eff_{\rm tot}^{\rm rel} $ are the products of the relative selection and PID efficiencies
discussed previously. The results for the branching fractions are

\begin{align*}
{\br(\btodkpipi)\over\br(\btodpipipi)} &= (5.9\pm1.1\pm0.5)\times10^{-2}, \nonumber \\ 
{\br(\btodzerokpipi)\over\br(\btodzeropipipi)} &= (9.4\pm 1.3\pm0.9)\times10^{-2}, \nonumber \\
\end{align*}

\noindent where the first uncertainties are statistical and the second are from the systematic sources
discussed below. 

Most systematic uncertainties cancel in the measured ratios of branching fractions; only those that
do not are discussed below.
One source of uncertainty comes from modeling of the $K^-\pi^+\pi^-$ final state. In Ref.~\cite{Aaij:2011rj},
we compared the $p$ and $p_{\rm T}$ spectra of $\pi^{\pm}$ from $X_d$, and they agreed well with simulation.
We have an insufficiently large data sample to make such a comparison in the CS signal decay modes. The
departure from unity of the efficiency ratios obtained from simulation are due to differences in the $p_{\rm T}$ 
spectra between the $X_d$ daughters in CF decays and the $X_s$ daughters in the CS decays. These differences
depend on the contributing resonances and the daughters' masses. We take the full difference of the relative 
efficiencies from unity (4.6\% for $\Bzb$ and 6.1\% for $B^-$) as a systematic uncertainty.

The kaon PID efficiency includes uncertainties from the limited size of the data set used for 
the efficiency determination, the limited number of events in the MC sample over which we average,
and possible systematic effects described below. The statistical precision is taken as the RMS width of 
the kaon PID efficiency distribution
obtained from pseudo-experiments, where in each one, the kaon PID efficiencies in each ($p$, $\eta$) bin 
are fluctuated about their nominal values within their uncertainties. This contributes 1.5\% to the overall 
kaon PID efficiency uncertainty. We also consider the systematic error in using the $D^*$ data
sample to determine the PID efficiency. The procedure is tested
by comparing the kaon PID efficiency using a MC-derived efficiency matrix with the efficiency obtained by 
directly requiring $\Delta{\rm ln}\mathcal{L}(K-\pi)>8$ on the kaon from $X_s$ in the signal MC. The relative difference is found to be 
$(3.6\pm1.9)\%$.
We take the full difference of 3.6\% as a potential systematic error. The total kaon PID uncertainty is 3.9\%.

The fit model uncertainty includes 3\% systematic uncertainty in the yields from the normalization modes~\cite{Aaij:2011rj}. 
The uncertainties in the CS signal fits are obtained by varying each of the signal shape parameters
within the uncertainty obtained from the CF mode data fits. The signal shape parameter uncertainties
are 2.7\% for $\Bzb$ and  2.5\% for $B^-$. For the specific $b$-hadron background shapes, we obtain the
uncertainty by refitting the data 100 times, where each fit is performed with all background shapes
fluctuated within their covariances and subsequently fixed in the fit to data (1\%). The uncertainties
in the yields from the assumed exponential shape for the combinatorial background are estimated by
taking the difference in yields between the nominal fit and one with a linear shape for the combinatorial
background (2\%). In total, the relative yields are uncertain by 4.5\% for $\Bzb$ and 4.4\% for $B^-$.

The limited number of MC events for determining the relative efficiencies contributes 4.1\% and 3.8\%
to the $\Bzb$ and $B^-$ branching fraction ratio uncertainties, respectively.
Other sources of uncertainty are negligible. 
In total, the uncertainties on the ratio of branching fractions are 8.6\% 
for $\Bzb$ and 9.3\% for $B^-$.

We have also looked at the substructures that contribute to the CS final states. Figure~\ref{fig:B2D0KPiPi_KPiPiPlots}
shows the observed distributions of (a) $K^-\pi^+\pi^-$ invariant mass, (b) $M(D^0\pi^+\pi^-)-M(D^0)$ invariant mass difference, 
(c) $K^-\pi^+$ invariant mass, and (d) $\pi^+\pi^-$ invariant mass for $\btodzerokpipi$. We show events in the $B$ mass
signal region, defined to have an invariant mass from 5226$-$5326~MeV/$c^2$, and events from the
high-mass sideband (5350$-$5550~MeV/$c^2$), scaled by the ratio of expected background yields
in the signal region relative to the sideband region. An excess of events is observed predominantly in the low 
$K^-\pi^+\pi^-$ mass region near 1300$-$1400~MeV/$c^2$, and the number of signal events decreases with increasing
mass. In Fig.~\ref{fig:B2D0KPiPi_KPiPiPlots}(b) there appears to be an excess of $\sim$10 events in the region 
around 550$-$600 MeV/$c^2$, which suggests contributions from $D_1(2420)^0$ or $D_2^*(2460)^0$ meson decays. 
These decays can also be used for measuring the weak phase $\gamma$~\cite{Sinha:2004ct}. This yield, relative to the total, 
is similar to what was observed in $\btodzeropipipi$ decays~\cite{Aaij:2011rj}.
Figures~\ref{fig:B2D0KPiPi_KPiPiPlots}(c) and (d) show significant enhancements at the $\bar{K}^{*0}$ and
$\rho^0$ masses, consistent with decays of excited strange states, such as the $K_1(1270)^-$, $K_1(1400)^-$ 
and $K^*(1410)^-$. Similar distributions are observed for the $\btodkpipi$, except that no excess of events is
observed near 550$-$600 MeV/$c^2$ in the $M(D^0\pi^+\pi^-)-M(D^0)$ invariant mass difference.

\begin{figure}[h]
\centering
\includegraphics[width=0.90\textwidth]{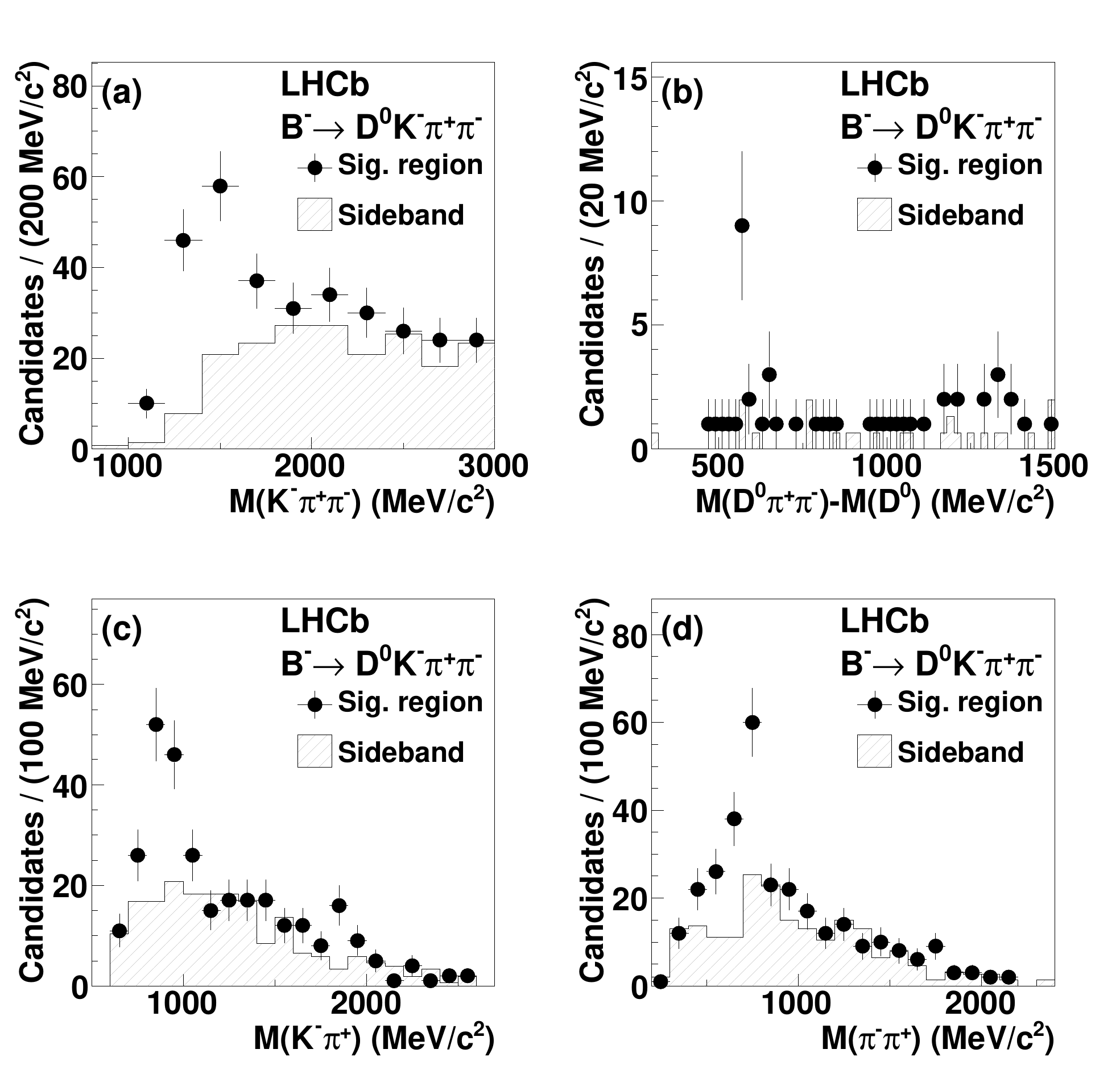}
\caption{Invariant masses within the $\btodzerokpipi$ system. Shown are (a)  $M(K^-\pi^+\pi^-)$ , 
(b)  $M(D\pi^+\pi^-)-M(D)$, (c) $M$($K^-\pi^+)$, and (d) $M(\pi^+\pi^-)$ . The points with error bars correspond to
the signal region, and the hatched histograms represent the scaled sideband region.}
\label{fig:B2D0KPiPi_KPiPiPlots}
\end{figure}

In summary, we report first observations of the Cabibbo-suppressed decay modes $\btodkpipi$ and
$\btodzerokpipi$ and measurements of their branching fractions relative to $\btodpipipi$ 
and $\btodzeropipipi$. The $\btodzerokpipi$ decay is particularly interesting because, with
more data, it can be used to measure the weak phase $\gamma$, using 
similar techniques as for $\btodzerok$ and $\btodzerokstar$.

\section*{Acknowledgements}

\noindent We express our gratitude to our colleagues in the CERN accelerator
departments for the excellent performance of the LHC. We thank the
technical and administrative staff at CERN and at the LHCb institutes,
and acknowledge support from the National Agencies: CAPES, CNPq,
FAPERJ and FINEP (Brazil); CERN; NSFC (China); CNRS/IN2P3 (France);
BMBF, DFG, HGF and MPG (Germany); SFI (Ireland); INFN (Italy); FOM and
NWO (The Netherlands); SCSR (Poland); ANCS (Romania); MinES of Russia and
Rosatom (Russia); MICINN, XuntaGal and GENCAT (Spain); SNSF and SER
(Switzerland); NAS Ukraine (Ukraine); STFC (United Kingdom); NSF
(USA). We also acknowledge the support received from the ERC under FP7
and the Region Auvergne.

\bibliographystyle{LHCb}
\bibliography{main}

\end{document}